\documentclass[aps,
reprint,twocolumn,showpacs,superscriptaddress,groupedaddress]{revtex4}  
\usepackage{graphicx}  
\usepackage{dcolumn}   
\usepackage{bm}        
\usepackage{amssymb}   
\usepackage{amsmath}
\usepackage{comment}
\usepackage{hyperref}

\begin{document}



\title{Spinor-Helicity Formalism for Massless Fields in AdS${}_4$}
\author{Balakrishnan Nagaraj}
\email{nbala@physics.tamu.edu}
\author{Dmitry Ponomarev}
\email{ponomarev@tamu.edu}
\affiliation{George P. and Cynthia W. Mitchell Institute for Fundamental Physics and Astronomy\\
Texas A\&M University, College Station, TX 77843-4242, USA}

\begin{abstract}
In this letter we suggest a natural  spinor-helicity formalism for massless fields in AdS${}_4$. It is based on the standard realization of the AdS${}_4$ isometry algebra $so(3,2)$ in terms of differential operators acting on $sl(2,\mathbb{C})$ spinor variables. We start by deriving the AdS counterpart of plane waves in flat space and then use them to evaluate simple scattering amplitudes. Finally, based on symmetry arguments we classify all three-point amplitudes involving massless spinning fields.  As in flat space, we find that the spinor-helicity formalism allows to construct additional consistent interactions compared to approaches employing Lorentz tensors.
\end{abstract}

\pacs{11.15.-q,11.25.Tq,11.80.Cr}
\maketitle


\section{\label{intro}Introduction}

The spinor-helicity formalism by now has established itself as the most efficient framework for representing on-shell scattering amplitudes of massless particles in the 4d Minkowski space, see e. g. \cite{Dixon:1996wi,Elvang:2013cua} for reviews. Success of this formalism in the original setup has motivated various extensions --- to other dimensions \cite{Cheung:2009dc,CaronHuot:2010rj,Boels:2012ie,Bandos:2016tsm,Bandos:2017eof}  and to massive particles \cite{Conde:2016vxs,Arkani-Hamed:2017jhn,Jha:2018hag}. At the same time, literature on the spinor-helicity formalism in curved space remains very limited.  In \cite{Maldacena:2011nz} a version of the spinor-helicity formalism  for massless fields in dS${}_4$ was proposed.  Despite its virtues, in some aspects it departs from the  spinor-helicity formalism in flat space, e.g. it loses manifest Lorentz covariance.

In this letter we make an alternative proposal for the spinor-helicity formalism in (A)dS${}_4$, which has all features of the flat space formalism and, in particular, reduces to the latter for conformal theories. In this respect, it is worth mentioning the twistor approach \cite{PR}, 
which is naturally adapted to describing fields in conformally flat spaces. Upon specializing to AdS backgrounds, the twistor approach can be used to obtain certain representations of massless scattering amplitudes in AdS${}_4$ \cite{Adamo:2012nn,Skinner:2013xp,Adamo:2013tja,Adamo:2015ina,Haehnel:2016mlb,Adamo:2016ple,Adamo:2018srx}. Our approach provides a different perspective on these results, as it allows to compute amplitudes directly from the space-time action
 and does not rely on a twistor-space description of the massless theories in question \footnote{A related approach that makes all higher spin symmetry manifest was developed in \cite{Colombo:2012jx,Didenko:2012tv,Gelfond:2013xt}.}.

Besides having obvious motivations --- e. g. development of tools that could facilitate  computations of holographic/inflationary correlators and simplify their analytic structure  ---  we  are also interested in gaining a better understanding of higher-spin interactions in flat and curved backgrounds and clarifying their relation. In particular, as was emphasized recently \cite{Bengtsson:2014qza,Conde:2016izb,Sleight:2016xqq}, in flat space  the spinor-helicity formalism and the light-cone approach admit additional cubic higher-spin vertices compared to those built of Lorentz tensors. Moreover, these additional vertices are  crucial for consistency of higher-spin interactions  \cite{Metsaev:1991mt,Metsaev:1991nb} and are present in chiral higher-spin theories \cite{Ponomarev:2016lrm,Ponomarev:2017nrr,Skvortsov:2018jea}, see also \cite{Devchand:1996gv} for a related earlier result. 
Until recently, the fate of additional interactions in AdS was not clear. 
In \cite{Metsaev:2018xip} the expectation that they also exist in AdS${}_4$ was confirmed in the light-cone approach. 
Below we classify all consistent 3-point amplitudes for massless particles in AdS${}_4$ using the spinor-helicity formalism and find agreement with \cite{Metsaev:2018xip}.

\section{Spinor-Helicity and Flat Space}
The basic fact about massless representations in the $4d$ Minkowski space is that they are labelled by two quantum numbers --- helicity $h$ and momentum $p$. Using the isomorphism $so(3,1)\sim sl(2,\mathbb{C})$  we have $p_\mu = -\frac{1}{2}(\sigma_\mu)^{\dot\alpha\alpha}\lambda_{\alpha}\bar\lambda_{\dot\alpha}$. For $h\ge 0$ the associated state can be represented by a potential
\begin{equation}
\label{2nov2n}
\varphi^{h}_{\nu_1\dots\nu_h}= \varepsilon^+_{\nu_1}\dots \varepsilon^+_{\nu_h}e^{ipx}, 
\end{equation}
where $\varepsilon^+_\nu$ is a polarization vector defined by
\begin{equation}
\label{2nov3n}
\varepsilon^+_\nu = \frac{(\sigma_\nu)^{\dot\alpha\alpha}\mu_{\alpha}\bar\lambda_{\dot\alpha}}{\mu^\beta\lambda_\beta}.
\end{equation}
Here $\mu$ is an auxiliary spinor and the ambiguity of its choice reflects gauge ambiguity. Alternatively, states can be represented by gauge invariant field strengths.  For (\ref{2nov2n}) the field strength reads
\begin{equation}
\label{2nov4n}
F^{h}_{\dot\alpha_1\dots\dot\alpha_{2h}} = \bar\lambda_{\dot\alpha_1}\dots \bar\lambda_{\dot\alpha_{2h}}e^{ipx}.
\end{equation}
 Extension to $h<0$ and to fermions is straightforward. Once plane wave solutions (\ref{2nov2n}) are available, 
 one can evaluate amplitudes using the Feynman rules in any theory of massless fields.
 
The amplitudes  are strongly constrained by Poincare covariance. These constraints allow to fix 3-point amplitudes up to a coupling constant \cite{Benincasa:2007xk} to be
\begin{equation}
\begin{split}
\label{1nov10}
{\cal A}_{\rm I}(h_1,h_2,h_3) &=[12]^{d_{12,3}}[23]^{d_{23,1}}[31]^{d_{31,2}} \delta^4(p),\\
{\cal A}_{\rm II}(h_1,h_2,h_3)& =\langle 12\rangle^{-d_{12,3}}\langle 23\rangle^{-d_{23,1}}\langle 31\rangle^{-d_{31,2}} \delta^4(p).
\end{split}
\end{equation}
Here $[ij]\equiv \epsilon_{\dot\alpha\dot\beta} \bar\lambda_i^{\dot\alpha}\bar\lambda_j^{\dot\beta}$ and $\langle ij \rangle \equiv \epsilon_{\alpha\beta}\lambda_1^{\alpha}\lambda_2^\beta$, 
$d_{ij,k}\equiv h_i+h_j-h_k$ and $p\equiv\sum_i p_i$ is the total momentum. To make (\ref{1nov10}) non-trivial one assumes that momenta are complex, hence $\lambda$ and $\bar\lambda$ are not complex conjugate to each other. Then, ${\cal A}_{\rm I}$ (${\cal A}_{\rm II}$) is singular for $\sum_i h_i<0$ ($\sum_i h_i>0$) in the limit of real momenta and should be dropped as physically irrelevant.

\section{AdS${}_4$ and Plane Waves}

Massless representations  of the AdS${}_4$  isometry algebra  $so(3,2)$ can be obtained by deforming the flat space translation generator as follows \footnote{In fact, by appropriately deforming the translation generator one can realize all symmetric representations of $so(3,2)$, $so(4,1)$ and $iso(3,1)$. This was explicitly done in the vector language in \cite{Ponomarev:2010st}. For further developments in the spinor language see \cite{JM}.}
\begin{equation}
\label{1nov12}
\mathcal{P}_{\alpha\dot{\alpha}} =\lambda_{\alpha}\bar{\lambda}_{\dot{\alpha}}-{R^{-2}} {\partial}/{\partial\lambda^{\alpha}}{\partial}/{\partial\bar{\lambda}^{\dot{\alpha}}},
\end{equation}
where $R$ is the AdS radius.
This realization of massless representations is often referred to as the twisted adjoint representation \cite{Bekaert:2005vh}. Similarly to what happens in flat space, all algebra generators commute with the helicity operator $2H\equiv\bar\lambda^{\dot\alpha}\bar\partial_{\dot\alpha}-\lambda^\alpha \partial_{\alpha}$, which allows to split the representation space into representations of definite helicity. 

For our further purposes it will be convenient to choose coordinates in AdS that make Lorentz symmetry manifest. Starting from the ambient space description of AdS as a hyperboloid  $X^MX_M=-R^2$, $M=0,1,\dots, 4$ and making the stereographic projection from $X^M=(0,\dots,0,-R)$, followed by the appropriate rescaling,
we arrive at intrinsic coordinates $x^\mu$, $\mu=0,1,2,3$ with the metric 
\begin{equation}
\label{28nov1}
ds^2 = \left(1-\frac{x^2}{4R^2} \right)^{-2} \eta_{\mu\nu}dx^\mu dx^\nu.
\end{equation}
The AdS boundary in these coordinates is given by $x^2=4R^2$, the patch $x^2<4R^2$ ($x^2>4R^2$)  corresponds to  $X^4>-R$ ($X^4>-R$) in ambient coordinates. We will refer to these patches as the inner and the outer patches, while their union will be referred to as the global AdS. Finally, we note that the inversion
$x^\mu  \leftrightarrow   x^\mu\, \frac{4R^2}{x^2}$
acts as the reflection with respect to the origin in  ambient space.

The AdS isometries act on bulk fields by Lie derivatives along Killing vectors. In our analysis Lorentz symmetry will be manifest, so we only specify Killing vectors associated with deformed translations. They act on scalar fields by
\begin{equation}
\label{1nov17}
    P_a=-i\left(1+\frac{x^2}{4R^2}\right)\delta^\mu_a\frac{\partial}{\partial x^\mu}+i\frac{x_a}{2R^2}x^\mu\frac{\partial}{\partial x^\mu}.
\end{equation}
To deal with spinning fields in terms of spinors we introduce a local Lorentz frame by means of the frame field
\begin{equation}
\label{28nov2}
    e^a_{\mu}=\left( 1-\frac{x^2}{4R^2}\right)^{-1}  \delta^a_{\mu},
    \end{equation}
$a=0,1,2,3$. It can be used to convert tensor fields from the coordinate basis to the local Lorentz basis, e.g. $A^a = e^a_\mu A^\mu$.
The frame field $e^a_{\mu}$ transforms as a 1-form with respect to diffeormorphisms. It is not hard to check that diffeomorphisms along (\ref{1nov17}) do not leave $e^a_{\mu}$ invariant. One can, however, complement  them with compensating local Lorentz transformations, so that the frame field becomes invariant. These compensating local Lorentz transformations then act on all fields carrying local Lorentz indices according to their index structure. In particular, for local Lorentz spinors we have
 \begin{equation}
\begin{split}
\label{1nov19}
    &(\delta_L P_{\alpha\dot{\alpha}}\cdot\bar{\lambda})^{\dot{\beta}}=\frac{i}{4R^2}\left(\delta^{\dot{\beta}}_{\dot{\alpha}}x_{\alpha\dot{\gamma}}+\epsilon^{\dot{\beta}\dot{\delta}}x_{\alpha\dot{\delta}}\epsilon_{\dot{\gamma}\dot{\alpha}}\right)\bar{\lambda}^{\dot{\gamma}},\\
    &(\delta_L P_{\alpha\dot{\alpha}}\cdot\lambda)^{\beta}=\frac{i}{4R^2}\left(\delta^{\beta}_{\alpha}x_{\gamma\dot{\alpha}}+\epsilon^{\beta\delta}x_{\delta\dot{\alpha}}\epsilon_{\gamma\alpha}\right)\lambda^{\gamma}.
\end{split}
\end{equation}
All spinor indices that we will encounter below refer to the local Lorentz basis.

Now we will find the AdS counterpart of flat plane wave solutions \footnote{This discussion parallels parts of \cite{Fronsdal:1974ew}.}.
As in flat space, this is necessary to give the amplitudes we are going to find later a familiar field-theoretic interpretation.
 The plane waves will be derived based on a consideration that they should serve as intertwining kernels between  two representations --- the spinor-helicity representation and the  space-time representation. We will focus on  plane waves for field strengths, as these are gauge invariant and do not require any auxiliary objects, such as reference spinors. Then, Lorentz invariance requires that indices of field strengths can only be carried by $\lambda_{\alpha}$, $\bar\lambda_{\dot\alpha}$, $(x\lambda)_{\dot\alpha}\equiv x_{\alpha\dot\alpha}\lambda^{\dot\alpha}$ or $(x\bar\lambda)_\alpha\equiv x_{\alpha\dot\alpha}\bar\lambda^{\dot\alpha}$. All the remaining spinor indices should be covariantly contracted, which implies that plane waves may also depend on  two scalars  $x_{\alpha\dot\alpha}\lambda^{\alpha}\bar\lambda^{\dot\alpha}$ and $x_{\alpha\dot\alpha}x^{\dot\alpha\alpha}$. Finally, we require that the action of the deformed translations on the plane wave in  representation (\ref{1nov12}) agrees with that in space-time (\ref{1nov17}), supplemented with compensating Lorentz transformations (\ref{1nov19}). This results in a differential equation that fixes the functional dependence of plane waves on $x_{\alpha\dot\alpha}\lambda^{\alpha}\bar\lambda^{\dot\alpha}$ and $x_{\alpha\dot\alpha}x^{\dot\alpha\alpha}$. 
For  $h\ge 0$ it has four linearly independent solutions
\begin{equation}
\begin{split}
\label{28nov3}
F^{\rm r|i}_{\dot\alpha_1\dots \dot\alpha_{2h}}&= \bar\lambda_{\dot\alpha_1}\dots \bar\lambda_{\dot \alpha_{2h}}\left(1-\frac{x^2}{4R^2}\right)_+^{1+h}e^{ipx},\\
F^{\rm r|o}_{\dot\alpha_1\dots \dot\alpha_{2h}}&= \bar\lambda_{\dot\alpha_1}\dots \bar\lambda_{\dot \alpha_{2h}}\left(1-\frac{x^2}{4R^2}\right)_-^{1+h}e^{ipx},\\
F^{\rm s|i}_{\alpha_1\dots\alpha_{2h}}
 &= \frac{(x\bar\lambda)_{\alpha_1}\dots  (x\bar\lambda)_{\alpha_{2h}}}{(x^2)^{{h}}}
\left(1-\frac{4R^2}{x^2}\right)_+^{1+{h}}e^{ipx \frac{4R^2}{x^2}},\\
F^{\rm s|o}_{\alpha_1\dots\alpha_{2h}}
 &= \frac{(x\bar\lambda)_{\alpha_1}\dots  (x\bar\lambda)_{\alpha_{2h}}}{(x^2)^{{h}}}\left(1-\frac{4R^2}{x^2}\right)_-^{1+h}e^{ipx \frac{4R^2}{x^2}},
\end{split}
\end{equation}
where $x_+\equiv x\theta(x)$ and $x_-\equiv -x\theta(-x)$.
Analogously, solutions can be constructed for $h<0$.

These solutions have the following properties. Plane waves $F^{\rm r|i}$  ($F^{\rm r|o}$) are supported on the inner (outer) patch that is
for $x^2<4R^2$ ($x^2>4R^2$).
The inversion   maps $F^{\rm r|i} \leftrightarrow F^{\rm s|o}$ and $F^{\rm s|i}\leftrightarrow F^{\rm r|o}$.  Solutions $F^{\rm s|i}$ ($F^{\rm s|o}$) are supported on $0<x^2<4R^2$ ($x^2<0$ and $x^2>4R^2$).
 One can also consider the following linear combinations \footnote{$F^{\rm r|g}$ was found in \cite{Bolotin:1999fa} by different methods. Implicitly AdS plane waves also appeared  in the twistor literature \cite{Hitchin:1980hp}.}
\begin{equation}
\begin{split}
\label{1nov20}
F^{\rm r|g}_{\dot\alpha_1\dots \dot\alpha_{2h}}&= \bar\lambda_{\dot\alpha_1}\dots \bar\lambda_{\dot \alpha_{2h}}\left(1-\frac{x^2}{4R^2}\right)^{1+{h}}e^{ipx},\\
F^{\rm s|g}_{\alpha_1\dots\alpha_{2h}}
 &= \frac{(x\bar\lambda)_{\alpha_1}\dots  (x\bar\lambda)_{\alpha_{2h}}}{(x^2)^{{h}}}
\left(1-\frac{4R^2}{x^2}\right)^{1+{h}}e^{ipx \frac{4R^2}{x^2}},
\end{split}
\end{equation}
which are supported on the global AdS patch. Note that both $F^{\rm r|g}$ and $F^{\rm r|i}$ reduce to familiar flat plane waves in the flat space limit $R\to\infty$. Referring to the behavior of solutions at $x\to 0$, we will call $F^{\rm r|g}$, $F^{\rm r|i}$ and $F^{\rm r|o}$ regular solutions, while $F^{\rm s|g}$, $F^{\rm s|i}$ and $F^{\rm s|o}$ will be called singular \footnote{Note that the dual representation to (\ref{1nov12}) differs by ${\cal P}\to -{\cal P}$, so the solutions for the massless module and its dual should be  both present in (\ref{28nov3}). Also note that both $F^{\rm r}$ and $F^{\rm s}$ behave as $z$ when $z\to 0$ in the Poincare coordinates, while $F^{\rm r}+F^{\rm s}$ behaves as $z^2$. Using the common holographic terminology, this allows to identify $F^{\rm r}+F^{\rm s}$ with normalizable modes. We leave investigation of the issue of non-normalizable modes for future research.}. 

At this point one may wonder whether splitting of the plane wave solutions into patches as in (\ref{28nov3}) is physically meaningful and whether it is enough to consider only solutions supported on the global patch (\ref{1nov20}). We do not have much to say about this except that splitting (\ref{28nov3}) is mathematically consistent with the symmetry arguments that we employed to derive these solutions. It is also worth remarking that for fermionic fields global solutions (\ref{1nov20}) feature square roots leading to ambiguities of the analytic continuation across the interfaces between the patches. Any such continuation is consistent with the symmetry arguments discussed above.

Finally, we would like to comment on the role of conformal symmetry in this discussion. Massless fields in 4d are conformally invariant \cite{Mack:1969dg
}, however, their description in terms of potentials breaks conformal invariance except for the spin one case. Given that AdS and flat spaces are conformally equivalent, this means that at least the regular solution in (\ref{1nov20}) could have been obtained by applying the  appropriate conformal transformation on a flat space plane wave solution.  Putting differently, our labelling of AdS plane waves  is consistent with the flat space one modulo conformal transformations. Conformal invariance also allows to conclude that the spin-1 potential is given by flat formula (\ref{2nov2n}). A thorough investigation of  potentials will be given elsewhere.

\section{AdS${}_4$ Scattering Amplitudes}

In AdS one can define tree-level scattering amplitudes as the classical action evaluated on the solutions to the linearized equations of motion. Below we will evaluate some simple amplitudes using plane wave solutions we have just obtained. We will focus on the scattering of regular plane waves, as they have smooth flat limit and a clearer connection to the familiar flat space amplitudes \footnote{For earlier discussions on the flat limit of higher-spin vertices see \cite{Boulanger:2008tg}.}. 

In the following we will encounter integrals \cite{GS}
\begin{widetext}
\begin{equation}
\begin{split}
\label{7nov4}
{\cal I}^{\rm r|i}_\lambda\equiv  \int d^4x{\left(1-\frac{x^2}{4R^2}\right)_+^{\lambda}} e^{i px}&=  2^{\lambda+6}{\Gamma(\lambda+1)}\pi iR^4
\left[ e^{- i \pi(\lambda-\frac{1}{2})} \frac{K_{\lambda+2}(-2iR(p^2+i0)^{\frac{1}{2}})}{(-2iR(p^2+i0)^{\frac{1}{2}})^{\lambda+2}} 
- \text{c.c.} \right],\\
{\cal I}^{\rm r|o}_\lambda\equiv  \int d^4x{\left(1-\frac{x^2}{4R^2}\right)_-^{\lambda} }e^{i px}&= 2^{\lambda+6} {\Gamma(\lambda+1)}\pi iR^4
\left[ e^{i\frac{\pi}{2}} \frac{K_{\lambda+2}(-2iR(p^2+i0)^{\frac{1}{2}})}{(-2iR(p^2+i0)^{\frac{1}{2}})^{\lambda+2}} 
- \text{c.c.} \right],
\end{split}
\end{equation}
\end{widetext}
where c.c. denote complex conjugate  and $K$ is the modified Bessel function of the second kind. These formulas should be understood in the sense of distributions and
are valid for  real $\lambda$ except negative integers, where ${\cal I}^{\rm r|i}_\lambda$ and ${\cal I}^{\rm r|o}_\lambda$ diverge.
In the following we will only  need ${\cal I}^{\rm r|i}_\lambda$ and ${\cal I}^{\rm r|o}_\lambda$  for integer values of $\lambda=n$.
We find it convenient to use the notation
\begin{equation}
\begin{split}
\label{7nov3}
{\cal I}^{\rm r|i}_{n} &=\left(1+\frac{\Box_p}{4R^2}\right)^{n}_+\delta^4(p), \quad
{\cal I}^{\rm r|o}_{n} =\left(1+\frac{\Box_p}{4R^2}\right)^{n}_-\delta^4(p),\\
{\cal I}^{\rm r|g}_{n}&={\cal I}^{\rm r|i}_{n}+(-1)^n {\cal I}^{\rm r|o}_{n}= \left(1+\frac{\Box_p}{4R^2}\right)^{n}\delta^4(p),
\end{split}
\end{equation}
which can be regarded as a result of a formal evaluation of the Fourier transform according to the rule $x^2\to -\Box_p$.
Representation (\ref{7nov3}) makes the distributional nature of amplitudes manifest and the flat space limit more intuitive.
Note that for non-negative $n$ the right hand side for ${\cal I}^{\rm r|g}_{n}$ in (\ref{7nov3}) is a well-defined distribution.
It can be shown that this result is consistent with representation (\ref{7nov4}), see \cite{GS}.

In these terms the $n$-point amplitudes for a scalar self-interaction vertex $L=\frac{1}{n!}\sqrt{-g} \varphi^n$
are given by
\begin{equation}
\label{7nov2}
{\cal A}^{\rm r|i}_n = {\cal I}^{\rm r|i}_{n-4}, \quad {\cal A}^{\rm r|o}_n = {\cal I}^{\rm r|o}_{n-4}, \quad {\cal A}^{\rm r|g}_n = {\cal I}^{\rm r|g}_{n-4}
\end{equation}
depending on the AdS patch we are using. For $n=3$ the amplitude is divergent, which is consistent with the standard  AdS/CFT analysis \cite{Freedman:1998tz}, where the 3-point Witten diagram for $\Delta=1$ scalars also gives a divergent result.

Similarly, we can evaluate more general vertices involving field strengths of spinning fields. For example, for 
$L=\frac{1}{2} \sqrt{-g}\varphi F^{\dot\alpha_1\dot\alpha_2}F_{\dot\alpha_1\dot\alpha_2}$
for different patches we find
\begin{equation}
\label{7nov6}
{\cal A}_3^{\rm r|i} = [ 23]^2  {\cal I}^{\rm r|i}_{1}, \quad
{\cal A}_3^{\rm r|o} = [ 23]^2  {\cal I}^{\rm r|o}_{1},\quad 
{\cal A}_3^{\rm r|g} = [ 23]^2  {\cal I}^{\rm r|g}_{1} .
\end{equation}
Amplitudes of the form ${\cal A}_3^{\rm r|g} $ have been previously derived in the twistor literature \cite{Adamo:2012nn,Skinner:2013xp,Adamo:2013tja,Adamo:2015ina,Haehnel:2016mlb,Adamo:2016ple,Adamo:2018srx}.

Finally, considering the Yang-Mills vertex, as a consequence of conformal invariance, we find exactly the same amplitude as in flat space, except that now we also have its  variants associated with different patches. In fact, conformal invariance of the Yang-Mills action implies that the same conclusion holds for all tree-level spinor-helicity amplitudes in AdS.

Having studied some simple examples, we will now move to the case of general spinning 3-point amplitudes. In contrast to the previous analysis, where we computed amplitudes using their field-theoretic definition, in the following, the amplitudes will be found by requiring correct transformation properties, that is solely from representation theory considerations.
As in flat space, Lorentz covariance is manifest and is achieved by combining spinors into spinor products. Moreover, once helicities on external lines are fixed, this imposes constraints on homogeneity degrees of spinors. For  amplitudes being genuine functions of spinor products this leads  to an ansatz
\begin{equation}
\label{7nov7}
{\cal A}(h_1,h_2,h_3) =[12]^{d_{12,3}}[23]^{d_{23,1}}[31]^{d_{31,2}} f(x,y,z),
\end{equation}
where
$x\equiv [12]\langle 12\rangle$, $y\equiv [23]\langle 23\rangle$ and  $z\equiv [31]\langle 31\rangle$. It only remains to impose correct transformation properties with respect to deformed translations 
\begin{equation}
\label{7nov8}
(\mathcal{P}^1_{\alpha\dot{\alpha}}+\mathcal{P}^2_{\alpha\dot{\alpha}}+\mathcal{P}^3_{\alpha\dot{\alpha}})\,{\cal A}(h_1,h_2,h_3)=0.
\end{equation}
This gives a  system of differential equations on $f(x,y,z)$. It can be shown that when at least one helicity is non-zero, one has four linearly independent solutions
 \footnote{What we actually derive  is the  solutions for $p^2>0$ and $p^2<0$, while their sewing  over $p^2=0$ is suggested by the examples that we previously considered.}
\begin{equation}
\begin{split}
\label{7nov10}
{\cal A}_{\rm I} &=[12]^{d_{12,3}}[23]^{d_{23,1}}[31]^{d_{31,2}}   {\cal I}^{\rm r|i}_{\sum h-1},\\
{\cal A}_{\rm II} &=[12]^{d_{12,3}}[23]^{d_{23,1}}[31]^{d_{31,2}}   {\cal I}^{\rm r|o}_{\sum h-1},\\
{\cal A}_{\rm III}& =\langle 12\rangle^{-d_{12,3}} \langle 23\rangle^{-d_{23,1}}\langle 31\rangle^{-d_{31,2}}   {\cal I}^{\rm r|i}_{-\sum h-1},\\
{\cal A}_{\rm IV} &=\langle 12\rangle^{-d_{12,3}} \langle 23\rangle^{-d_{23,1}}\langle 31\rangle^{-d_{31,2}}   {\cal I}^{\rm r|o}_{-\sum h-1},
\end{split}
\end{equation}
where ${\cal I}$'s are given by (\ref{7nov4}).
When all helicities are vanishing,  $f_{\rm I}$ coincides with $f_{\rm III}$ and $f_{\rm II}$ coincides with $f_{\rm IV}$.

Classification (\ref{7nov10}) is different from (\ref{1nov10}) only in two respects. First is that $so(3,2)$ covariance turns out to be consistent with splitting the global AdS into two patches, each being associated with its own amplitude. This explains why we get four solutions in (\ref{7nov10}) instead of two solutions in flat space. The second difference is that the flat space momentum-conserving delta functions in AdS are replaced with one of ${\cal I}$'s (\ref{7nov4}) depending on the patch one is interested in.
Based on the flat limit, where 
 ${\cal A}_{\rm I}$ and ${\cal A}_{\rm II}$ (${\cal A}_{\rm III}$ and ${\cal A}_{\rm IV}$) for $\sum_i h_i<0$ ($\sum_i h_i>0$) are singular for real momenta 
we argue that they should also be dropped in AdS as physically irrelevant.  It is worth mentioning that these amplitudes are divergent, see discussion below (\ref{7nov4}). The same refers to all amplitudes with $\sum_i h_i=0$.
Finally, we remark that ${\cal A}_{\rm I}$ (${\cal A}_{\rm II}$) for $\sum h=1$ ($\sum h =-1$) in flat space (\ref{1nov10}) are conformally invariant \cite{MPHD}. This explains why these are equal to the associated amplitudes in global AdS (\ref{7nov10}).

Amplitudes with three singular plane waves using the inversion reduce to amplitudes where all plane waves are regular. Amplitudes where regular and singular plane waves are mixed require separate analysis. If these are genuine functions, they should be given by linear combinations of (\ref{7nov10}). Another potential possibility is that they are given by distributions. In this respect it is worth noting that by considering an ansatz for a distribution supported on $p=0$ and requiring (\ref{7nov8}), we again end up with (\ref{7nov10}), where ${\cal I}$'s appear in representation (\ref{7nov3}).

\section{Conclusions}
In the present letter we suggested a natural generalization of the spinor-helicity formalism to (A)dS${}_4$. We started by generalizing the familiar flat space plane wave solutions to AdS and then used them to evaluate some simple 3-point amplitudes. We also classified all consistent spinning 3-point amplitudes by requiring correct transformation properties. We found that, as in flat space, for three generic spins, by picking different signs of helicities, one can construct four different parity-invariant amplitudes. At the same time, approaches that involve Lorentz tensors result only in two consistent parity-invariant structures both on the bulk \cite{Fradkin:1986qy,Vasilev:2011xf,Joung:2011ww,Boulanger:2012dx,Sleight:2016dba,Francia:2016weg} and boundary \cite{Giombi:2011rz,Zhiboedov:2012bm} sides.  This phenomenon directly generalizes an analogous one in flat space and is consistent with a recent analysis in the light-cone gauge \cite{Metsaev:2018xip}. 

The amplitudes that we computed were defined as the classical action evaluated on the particular basis of solutions to the linearized equations of motion. This definition is related to the holographic one --- where amplitudes are identified with boundary correlators and computed in the  bulk by Witten diagrams \cite{Maldacena:1997re,Gubser:1998bc,Witten:1998qj}  --- by a mere change of a basis for the states appearing on external lines.
Unlike bulk-to-boundary propagators, the plane wave solutions that we employed do not have a boundary limit that would allow to associate them with localized boundary sources. Instead, they have a transparent flat limit, which also makes the flat limit of the spinor-helicity amplitudes more intuitive. In this respect, our plane waves serve as the properly focused scattering states necessary to access flat space physics from holography, see e. g. \cite{Gary:2009ae,Heemskerk:2009pn,Maldacena:2015iua}. An explicit transformation relating the two bases will be given elsewhere.

An obvious future direction is to extend these results to higher-point amplitudes and see how various bulk scattering processes manifests themselves in amplitudes' analytic structure, see e. g. \cite{Penedones:2010ue,Fitzpatrick:2011hu,Arkani-Hamed:2018kmz} for related work.
Optimistically, clear understanding of the analytic structure of AdS spinor-helicity amplitudes may lead to the development of the on-shell methods, which are as efficient as in flat space.

Finally, our construction may be useful in shedding the light on how higher-spin no-go theorems, see \cite{Bekaert:2010hw} for review, can be circumvented in flat space. Thus far it is known how to construct higher-spin theories in flat space only in the chiral sector \cite{Ponomarev:2016lrm,Ponomarev:2017nrr,Skvortsov:2018jea}, while their parity-invariant completions are obstructed by non-localities. At the same time, higher-spin theories in AdS have solid support from holography \cite{Sezgin:2002rt,Klebanov:2002ja}. We believe that  the connection between higher spin theories in flat  and in AdS spaces does exist and both sides will benefit from its clarification.

\begin{acknowledgments}
We are grateful to V.~Didenko, R.~Metsaev, K.~Mkrtchyan, T.~McLoughlin and E.~Skvortsov for fruitful discussions and to T.~Adamo, E.~Skvortsov and A.~Tseytlin for comments on the draft. We would also like to thank E.~Joung for pointing \cite{Fronsdal:1974ew} to us as well as its relevance in the holographic context. 
The work of BN was supported by the Mitchell/Heep Chair in High Energy Physics, Texas A\&M University.
The work of DP was supported by NSF grants PHY-1521099 and PHY-1803875.
\end{acknowledgments}



\bibliography{sh.bib}
\bibliographystyle{utphys}

\end{document}